\documentstyle[12pt]{article}
\def\a{\alpha'}
\def\R{{\mathcal R}}
\def\T{{\mathcal T}}
\newcommand{\be}{\begin{equation}}
\newcommand{\ee}{\end{equation}}
\newcommand{\bea}{\begin{eqnarray}}
\newcommand{\eea}{\end{eqnarray}}
\begin{document}
\begin{titlepage}
\begin{center}
\hfill \texttt{CPHT-RR-046-0805}

\hfill \texttt{SPHT-T05/50}\\

\vskip 20mm

{\Huge Perturbation theory and stability analysis for
string-corrected black holes in arbitrary dimensions}

\vskip 10mm

Filipe Moura

\vskip 4mm

{\em Centre de Physique Th\'eorique, \'Ecole Polyt\'echnique\\
F91128 Palaiseau Cedex, France\\ and \\ Service de Physique
Th\'eorique, Orme des Merisiers, CEA/Saclay\\ F91191
Gif-sur-Yvette Cedex, France}\footnote{Present address: Instituto Superior T{\'e}cnico, 
Departamento de Matem{\'a}tica, Av. Rovisco Pais, 1049-001 Lisboa, Portugal.}

{\tt fmoura@spht.saclay.cea.fr}

\vskip 6mm
\end{center}
\vskip .2in

\begin{center} {\bf Abstract } \end{center}
\begin{quotation}
\noindent We develop the perturbation theory for $\R^2$
string-corrected black hole solutions in $d$ dimensions. After
having obtained the master equation and the $\a$-corrected
potential under tensorial perturbations of the metric, we study
the stability of the Callan, Myers and Perry solution under these
perturbations.
\end{quotation}
\vfill
\end{titlepage}
\eject
\setcounter{equation}{0}
String theory low energy effective actions have three different
types of contributions, with different origins. The classical terms
come from the expansion in $\a$ (world-sheet loops). The quantum
terms depend on the string coupling constant $g_s=\mbox{e}^{\phi}$;
they can be perturbative (coming from space-time loops) and
non-perturbative. In this work we consider only the classical $\a$ 
corrections, neglecting any kind of string quantum correction. 
Both the bosonic and the heterotic string theories have
corrections already at the first order in $\a$, which are at most
quadratic in the Riemann tensor. In these corrections we neglect the 
Ricci terms, which would only contribute in a higher order in $\a$; 
we are only considering an effective action which is perturbative in 
$\a$. All these theories also have antisymmetric
tensors in their massless spectra, which can always be
consistently set to zero. That will be the case in the
$\a$-corrected black hole solution we use in this work. Although
these theories lie respectively in 26 or 10 space-time dimensions,
we will consider in this article black hole space-times in generic
$d$-dimensions. This way we take, as our effective action in the
Einstein frame \cite{cmp89},
\be
\frac{1}{2 \kappa^2} \int \sqrt{-g} \left[R-\frac{4}{d-2}
\left(\partial^\mu \phi \right)
\partial_\mu \phi + \mbox{e}^{\frac{4}{2-d} \phi}
\frac{\lambda}{2} R^{\mu \nu \rho \sigma} R_{\mu \nu \rho
\sigma} \right] \mbox{d}^dx +\mbox{fermion terms,} \label{eef} \ee
with $\lambda=\frac{\a}{2}, \frac{\a}{4}, 0$ for bosonic,
heterotic and superstrings, respectively.

The corrected bosonic equations of motion for the dilaton and the
graviton are, to this order, \bea \nabla^2 \phi -
\frac{\lambda}{4} \mbox{e}^{\frac{4}{2-d} \phi} \left(R_{\rho
\sigma \lambda \tau} R^{\rho \sigma \lambda \tau} \right) &=& 0,
\label{bdfe} \\ R_{\mu \nu}+ \lambda \mbox{e}^{\frac{4}{2-d}
\phi} \left(R_{\mu \rho \sigma \tau} R_\nu^{\ \rho \sigma \tau}
- \frac{1}{2(d-2)} g_{\mu \nu} R_{\rho \sigma \lambda \tau}
R^{\rho \sigma \lambda \tau} \right) &=& 0. \label{bgfe} \eea


We are interested in studying the behaviour of a string-corrected
black hole solution under perturbations. We will be studying these
perturbations in generic $d$ spacetime dimensions \cite{kodama-ishibashi-0}, taking as background metric 
\be
d\,s^2=-f(r)\,d\,t^2 + f^{-1}(r)\,d\,r^2 +r^2\,d\,\Omega^2_{d-2}
\label{schwarz} \ee
with $d\Omega_{d-2}^2=\gamma_{ij} \left(\theta\right)\,d \theta^i\,d
\theta^j$, $\gamma_{ij}=g_{ij}/r^2$ being the metric of a $(d-2)$-sphere $S^{d-2}$.

One can in general consider perturbations to the metric and any
other physical field of the system under consideration. General tensors of rank at most 2 on the $(d-2)$-sphere can be
uniquely decomposed in their scalar, vectorial and (for $d>4$)
tensorial components. In this work we only consider tensorial (in $S^{d-2}$)
perturbations to the metric, given by $h_{\mu\nu}=\delta
g_{\mu\nu}$ (as we will show, we can consistently
set the tensorial perturbation to the dilaton to 0). These
perturbations are worked out in \cite{kodama-ishibashi-2}, where
it is shown that they can be written as
\be
h_{ij}=2 r^2\left(y^a\right) H_T\left(y^a\right)
\T_{ij}\left(\theta^i \right), \, h_{ir}= h_{it}=0, h_{rr}= h_{tr}=h_{tt}=0
\label{htensor} \ee with $\T_{ij}$ satisfying \be
\left(\gamma^{kl} D_k D_l + k_T \right) \T_{ij}= 0, \, D^i
\T_{ij}=0, \, g^{ij} \T_{ij}=0. \label{propt} \ee $D_i$ is the $S^{d-2}$ covariant derivative; $\T_{ij}$ are the eigentensors of the $S^{d-2}$ laplacian; on the same reference \cite{kodama-ishibashi-2}, it is
also shown that the eigenvalues are given by
$-k_T=2-l\left(l+d-3\right), l=2, 3, 4\ldots$

We actually need the variation of the components of the Riemann tensor. Using the components of $h_{\mu \nu}$ given in
(\ref{htensor}) and the Palatini equation, one gets \bea \delta R_{ijkl}&=& \left[\left(2f
-1 \right)H_T + f
\partial_r H_T \right] \left(g_{il} \T_{jk} - g_{ik}
\T_{jl} - g_{jl} \T_{ik} + g_{jk} \T_{il} \right) \nonumber
\\ &+&r^2 H_T \left(D_i D_l \T_{jk} - D_i
D_k \T_{jl} - D_j D_l \T_{ik} + D_j D_k \T_{il} \right)
\label{drtensori} \\ \delta R_{itjt}&=& \left[-r^2 \partial_t^2
H_T + \frac{1}{2} f f' r^2 \partial_r H_T + f f' r H_T \right]
\T_{ij} \\ \delta R_{irjr}&=& \left[-r \frac{f'}{f} H_T -
-\frac{1}{2} r^2 \frac{f'}{f}
\partial_r H_T -2r \partial_r H_T -r^2 \partial^2_r H_T \right]
\T_{ij}
\\ \delta R_{abcd}&=&0. \label{drtensora} \eea
Using the explicit form of the Riemann tensor and the variations (\ref{htensor}) and
(\ref{drtensori}-\ref{drtensora}), one can perturb the field equations (\ref{bdfe}) and (\ref{bgfe}). 
From (\ref{bdfe}), we are able to show that we can consistently set the dilaton perturbation $\delta
\phi=0$. By perturbing (\ref{bgfe}) we are able to determine the equation for $H_T,$ which is given by
\bea 
&&\left(1-2 \lambda \frac{f'}{r} \right) \left(\partial^2_t H_T 
-f^2 \partial^2_r H_T \right) \nonumber \\
&+& \left[(2-d) \frac{f^2}{r} - ff' 
+\lambda \left(4 (4-d) \frac{f^2}{r^3}\left(1-f\right) 
+ 4 \frac{f^2}{r^2} f' +2\frac{f}{r} f'^2 \right) \right] 
\partial_r H_T \nonumber \\
&+& \left[k_T \frac{f}{r^2} \left(1+ 
\frac{4 \lambda}{r^2} \left(1-f \right) \right) - 2 \frac{ff'}{r} 
+ 2(d-2)\frac{f}{r^2} - 2(d-3) \frac{f^2}{r^2} \right. \nonumber \\ 
&+&\left. \lambda \frac{f}{r^2} 
\left(\frac{2+2d}{r^2} -4(d-1) \frac{f}{r^2}+2(d-3) \frac{f^2}{r^2}
- \frac{\left( f''\right)^2}{d-2} r^2\right) \right] H_T =0.
\label{master0} \eea
We now write the equation
above in the form of a master equation \be \frac{\partial^2
\Phi}{\partial r_\ast^2} - \frac{\partial^2 \Phi}{\partial t^2} =:
V_T \Phi. \label{ikmaster} \ee For that, first we write the
perturbation equation in terms of the tortoise coordinate
$r_\ast$, defined by $d r^\ast/d r=1/f$. As carefully explained
in \cite{Moura:2006pz}, following the procedure introduced in
\cite{Dotti:2005sq} we derive our master function and potential:
\bea \Phi&=&
\frac{H_T}{\sqrt{f}} \exp\left(\int \frac{\frac{f'}{f}
+\frac{d-2}{r} + \frac{4}{r^3} (d-4) \lambda (1-f) - \frac{4}{r^2}
\lambda f' - \frac{2}{r f} \lambda f'^2}{2-\frac{4}{r}\lambda f'}
dr\right) \nonumber \\ 
V_T (f)&=& \left(\frac{1}{1-2 \lambda \frac{f'}{r}}\right)^2  
\left(1+ \frac{4 \lambda}{r^2} \left(1-f \right) \right) \left[
\frac{d-4}{4r^2} \left(1+ \frac{4 \lambda}{r^2} \left(1-f \right) \right) + 
\frac{2 \lambda f'' -1}{2r^2} \right] \nonumber \\
&+& \frac{1}{1-2 \lambda \frac{f'}{r}} \left[\left(k_T +2\right) 
\frac{f}{r^2} + 2 (d-3) \frac{f (1-f)}{r^2} + \frac{d-8}{2} 
\frac{f f'}{r} - \frac{\lambda}{d-2} f \left( f''\right)^2 \right. 
\nonumber \\
&+& \left. 4 \lambda  \left(k_T + 2 \right) \frac{f (1-f)}{r^4} +
2(d-3) \lambda \frac{f (1-f)^2}{r^4}+ 2 (d-4) \lambda \frac{f(1-f) f'}{r^3} 
\right] \nonumber \\
&+& \frac{f f'}{r} + (d-4) \frac{f^2}{r^2}.
\eea 

This is the potential for tensor perturbations of any kind of
$R^2$-corrected black holes in $d$ dimensions, in terms of which
the perturbation equation (\ref{master0}) is written as a ``master
equation'' like (\ref{ikmaster}).

To study the stability of a solution, we use the ``S-deformation
approach'' first introduced in \cite{kodama-ishibashi-2} and
developed in \cite{Dotti:2005sq}. After having obtained the
potential $V_T (f)$, we assume that its solutions are of the form
$\Phi(r_*,t)=e^{i\omega t} \phi(r_*),$ such that
$\partial\Phi/\partial t= i\omega \Phi.$ The master equation is
then written in the Schr\"odinger form $A \Phi= \omega^2 \Phi$,
and a solution to the field equation is then stable if the
operator $A$ is positive definite with respect to the following
inner product: $$\left\langle \phi, A \phi \right \rangle =
\int_{-\infty}^{+\infty} \overline{\phi} (r_*) \left[
-\frac{d^2}{dr_*^2} +V \right] \phi(r_*) \ dr_* =
\int_{-\infty}^{+\infty} \left[ \left|\frac{d \phi}{dr_*}\right|^2
+ V \left|\phi \right|^2 \right]\ dr_*.$$ Defining 
$D=\frac{d}{dr_*}-\frac{f H_T}{\Phi} \frac{d}{dr} \left( 
\frac{\Phi}{H_T}\right)$ and after some algebraic
tricks \cite{Moura:2006pz}, we are left with $$\left\langle \phi, A \phi \right
\rangle = \int_{-\infty}^{+\infty} \left|D \phi \right|^2 \ dr_* +
\int_{-\infty}^{+\infty} \frac{Q}{f} \left|\phi \right|^2 \
dr_*,$$ with 
\bea
\frac{Q}{f}&=& \frac{1}{1-2 \lambda \frac{f'}{r}} \frac{1}{r^2} \left[\left(k_T +2\right)
\left(1+ \frac{4 \lambda}{r^2} \left(1-f \right) \right) + (2d-6) (1-f) 
\left(1+ \frac{\lambda}{r^2} \left(1-f \right) \right) \right. 
\nonumber \\ &-& \left. 2r f'
- \frac{\lambda}{d-2} \left( f''\right)^2 r^2 \right] \label{qf}
\eea
All that is necessary to guarantee the stability 
is to check the positivity of $\frac{Q}{f}.$

We considered the
$R^2$-corrected black hole solution of the type of (\ref{schwarz})
studied in \cite{cmp89}. Its only free parameter is $\mu$,
which is related to the classical ADM black hole mass through
$m_{cl}= \frac{\left(d-2\right) A_{d-2}}{\kappa^2} \mu, A_n$ being
the area of the unit $n-$sphere.

For the classical Schwarzschild-Myers-Perry solution, we have
$f(r)=1-\frac{2 \mu}{r^{d-3}}$. In order to introduce the
$\a$-corrections to this solution, we choose a coordinate system
in which the position of the horizon, given by $r= \left(2
\mu\right)^{\frac{1}{d-3}}=: r_H,$ is not changed. According to
\cite{cmp89} $f(r)$ is given, in this coordinate system, by \be
f(r)=\left(1-\left(\frac{r_H}{r}\right)^{d-3} \right) \left[1-
\lambda \frac{(d-3)(d-4)}{2} \frac{r^{d-5}_H}{r^{d-1}}
\frac{r^{d-1}-r_H^{d-1}}{r^{d-3} - r^{d-3}_H} \right]. \label{fr2}
\ee


This solution has as free parameters the
inverse string slope $\lambda$, the black hole mass parameter
$\mu$ (or, equivalently, the horizon radius $r_H$) and the
spacetime dimension $d$. Since $\lambda$ is a perturbative
parameter, we should take it small (say $\lambda \ll 1$), for the
potential to make sense. For small values of $\lambda$, for each
value of $d$ between 5 and 10, and for a wide range of values of
$\mu$, we have studied numerically and made plots of $\frac{Q}{f}$
as it is given by (\ref{qf}), and we always found positive values.
From this numerical study we conclude that this solution is stable under 
tensor perturbations for every relevant
spacetime dimension, for every value of the black hole mass.

Given the potential for the metric tensorial perturbations, we 
have also obtained an analytical proof of the stability and computed the 
absorption cross section of the black hole given by (\ref{fr2}). 
For more details see \cite{Moura:2006pz}. 

\paragraph{Acknowledgements}
\noindent
This work has been supported by a Chateaubriand
scholarship from EGIDE and by fellowship BPD/14064/2003 from
Funda\c c\~ao para a Ci\^encia e a Tecnologia, and is part of a 
joint project with Ricardo Schiappa \cite{Moura:2006pz}.

\end{document}